\pdfoutput=1

\documentclass[12pt]{article}
\usepackage{amsmath,amssymb,latexsym,cite,eucal,theorem,
    fancyvrb,enumerate}
\usepackage[matrix,arrow,frame,import,curve,color]{xy}
\usepackage{tikz}
\usetikzlibrary{calc}
\usetikzlibrary{arrows}

\usepackage{graphicx}

\newtheorem{prop}{Proposition}

\newtheorem{definition}{Definition}
{\theorembodyfont{\rmfamily}}

\newif\iffigs\figstrue

\DeclareFontFamily{U}{rsf}{}
\DeclareFontShape{U}{rsf}{m}{n}{
  <5> <6> rsfs5 <7> <8> <9> rsfs7 <10-> rsfs10}{}
\DeclareMathAlphabet\Scr{U}{rsf}{m}{n}

\def\O{\Scr{O}}
\def\C{{\mathbb C}}
\def\P{{\mathbb P}}

\def\R{{\mathbb R}}
\def\Z{{\mathbb Z}}

\def\End{\operatorname{End}}

\def\coker{\operatorname{coker}}

\def\Spec{\operatorname{Spec}}

\def\GU{\operatorname{U{}}}

\def\id{{\mathbf{1}}}

\newcommand{\Moverbar}[1]{\mkern
  7mu\overline{\mkern-7mu#1\mkern+2.0mu}\mkern-2mu}

\def\CY{Calabi--Yau}

\def\GLSM{gauged linear $\sigma$-model}

\def\cM{{\Scr M}}
\def\cMcp{\Moverbar{\cM}}

\def\cA{{\Scr A}}

\def\cE{{\Scr E}}

\def\DC{\mathbf{D}^{\mathrm{b}}}

\def\ff#1#2{{\textstyle\frac{#1}{#2}}}

\def\eqn#1#2{\begin{equation}#2
  \ifx{#1}{}\else\label{#1}\fi\end{equation}}

\begin{document}

\begin{titlepage}
\begin{flushright}
November 2014
\end{flushright}
\vspace{.5cm}
\begin{center}
\baselineskip=16pt
{\fontfamily{ptm}\selectfont\bfseries\huge
Exoflops in Two Dimensions\\[20mm]}
{\bf\large  Paul S.~Aspinwall
 } \\[7mm]

{\small

Department of Mathematics\\ Duke University, 
 Durham, NC 27708-0223 \\ \vspace{6pt}

 }

\end{center}

\begin{center}
{\bf Abstract}
\end{center}
An exoflop occurs in the gauged linear $\sigma$-model by varying the
K\"ahler form so that a subspace appears to shrink to a point and then
reemerge ``outside'' the original manifold. This occurs for K3
surfaces where a rational curve is ``flopped'' from inside to outside
the K3 surface. We see that whether a rational curve contracts to an
orbifold phase or an exoflop depends on whether this curve is a line
or conic. We study how the D-brane category of the smooth K3 surface
is described by the exoflop and, in particular, find the location of a
massless D-brane in the exoflop limit. We relate exoflops to
noncommutative resolutions.
\vfil\noindent

\end{titlepage}

\vfil\break


\section{Introduction}    \label{s:intro}

The venerable gauged linear $\sigma$-model \cite{W:phase} has been
used in a wide variety of applications in the past 20 years. The key
notion is that there is some K\"ahler moduli space that can be divided
up into ``phases'' and these phases are asymptotically the cones in
some secondary fan. Typical familiar and well-studied phases include
large radius limits, orbifold and Landau--Ginzburg models. These are
misleading, however, since they are not singular.\footnote{The large
  radius limit is, of course, singular at infinity but this is
  infinite distance in the moduli space and can be viewed in a
  controlled way as a decompactification. Our use of the term
  ``singular'' will mean at finite distance.} In a typical example
with $h^{1,1}\gg1$, one has a multitude of ``exoflops'' and ``bad
hybrids'' \cite{me:hybridm,Bertolini:2013xga} as singular phase
limits. In this paper we study this somewhat neglected exoflop
phase. In particular we will study them in the simplest setting they
appear, namely dimension two.

An exoflop occurs when an algebraic subset is contracted by deforming
the K\"ahler form. As one continues this deformation beyond the point
of contraction, a new component of the gauged linear $\sigma$-model
vacuum grows out external to the original space. The new component
intersects the original space at a singularity resulting from the
contraction. Only rational curves can be contracted in a K3 surface
but we will see that their embedding in the ambient toric variety will
determine whether an orbifold phase or an exoflop phase occurs on
their contraction.

It is well-known that singular theories lurk in the walls dividing the
phases from each other.  It might appear, then, that there are two
distinct sets of theories of particular interest:
\begin{itemize}
\item The phase limits that appear in the deep limit of each
  maximal cone in the secondary fan.
\item The singular theories in the wall separating these limits.
\end{itemize}
There really is no such intrinsic dichotomy as has been noted before
\cite{GMV:Ht,AdAs:masscat}. In this paper we will compare how a
singular theory can manifest itself dually as living in a wall or
living in a limit in the simplest case. That is, we consider
contracting a curve in a K3 surface such that the $B$-field is zero on
this curve. In the past this has generally been considered as the
singular theory that separates the large radius phase of the K3 surface
from some orbifold phase. Here we will also manifest it as an exoflop
limit in a gauged linear $\sigma$-model.

Naturally the intrinsic geometry, as viewed by a {\em non-linear\/}
$\sigma$-model, of these two possibilities is
identical and thus any associated physics in the string
compactification would be identical. However, the presentation of the
theories is quite different. In particular, the description of the
D-branes will be not at all the same and we focus on this below.

The D-branes in an orbifold phase include fractional
branes stuck at the orbifold point \cite{DM:qiv}. These ``resolve''
the category in the sense that they provide the objects in the derived
category that appear when the orbifold point is blown up according to
the McKay correspondence. In contrast to this we will see that these
extra D-branes in an exoflop appear as matrix factorizations living at
the North pole on the $\P^1$ sticking out of the singularity. This is
more in the flavour of noncommutative resolutions as we discuss below.

In section \ref{s:K3eg} we discuss the general difference between
lines and conics as they are contracted in a K3 surface. We use
the elegant technology of spherical functors to analyze the monodromy
and thus D-brane behaviour. In section \ref{s:geom} we analyze
two examples in detail to see how the D-branes understood geometrically in
the smooth phase appear in the exoflop. In the second example we have
a ``double exoflop'' where are chain of two $\P^1$'s are pushed out of
the K3 surface.


\section{Lines vs Conics}  \label{s:K3eg}

Consider a smooth curve $C$ of degree $m$ in $\P^2$. The case $m=1$ is a
line and the case $m=2$ is a conic. They are both isomorphic to
$\P^1$. We wish to extend this notion to more complicated embeddings
$C\subset V$, where $V$ can be, for example, a toric variety. 

\begin{definition}  \label{def:conic}
Let $C\cong\P^1$ and consider an embedding $i:C\to V$, where $V$ is a
toric variety. Let $m$ be the smallest positive integer such that
there exists a line bundle $\cE$ on $V$ with $i^*\cE=\O_C(m)$. If
$m=1$ we call $C$ a line and if $m=2$ we call $C$ a conic.
\end{definition}

\subsection{The Toric Construction} \label{ss:basic}

Let us quickly review the toric machinery we require. We refer to
section 2 of \cite{AdAs:masscat} for details of the construction
rather than copying it verbatim here. Here is a quick list of
notation:
\begin{itemize}
\item $N$ is a lattice of rank $d$ with dual lattice $M$.
\item $\cA$ is a collection of $n$ points lying in a hypersurface in
  $N$.
\item $\Sigma$ is a regular triangulation of $\cA$ or, depending on
  context, a fan over that triangulation.
\item $S=\C[x_1,\ldots,x_n]$ is the associated homogeneous coordinate
  ring and $B_\Sigma\subset S$ is the irrelevant ideal. $S$ has
  an $r$-fold grading (the $\GU(1)^r$-charges) , where $r=n-d$.
\item $Z_\Sigma=(\C^n-V(B_\Sigma)/(\C^*)^r$ is the associated
  toric variety (or stack).
\item $W$ is the superpotential, an element of $S$ invariant under
  $(\C^*)^r$.
\item $X_\Sigma$ is the critical point set (or stack) of $W$ in $Z_\Sigma$.
\end{itemize}

The {\em secondary polytope\/} has vertices corresponding to regular
triangulations of $\cA$ \cite{GKZ:book}. There is a dual {\em
  secondary fan.} Let $\cMcp$ denote the
corresponding toric variety. This is viewed as a natural
compactification of the ``moduli space of complexified K\"ahler
forms'' and is of dimension $r$.

Each regular triangulation of $\cA$ corresponds to a phase and
corresponds to a point in $\cMcp$. This point is called the ``phase
limit''.  The discriminant $\Delta\subset\cMcp$ is called the
``principal $A$-determinant'' in \cite{GKZ:book}. Singular conformal
field theories are contained in $\Delta$ and in the toric divisors in
$\cMcp$.

A one-dimensional edge of the secondary polytope corresponds to a
toric rational curve in $\cMcp$ ``joining'' two phase limits. This is
associated to a perestroika (see section 7.2.C of
\cite{GKZ:book}). This rational curve will intersect $\Delta$ at one
point \cite{AdAs:masscat}. Thus, this rational curve has three
interesting points --- the two phase limits and the intersection with
$\Delta$.

\subsection{Line Examples}  \label{ss:line}   

The simplest construction to see a line $C$ in a \CY\ surface is
non-compact (i.e., no superpotential). In this case $d=2$, $\cA$ is
three collinear equally spaced points and $S=\C[x_0,x_1,x_2]$ with
degrees $(-2,1,1)$.

There are two phases --- either we use the middle point (associated to
$x_0$) or we ignore it. If we use it then we have the large radius
phase with $Z$ the total space of $\O_C(-2)$. Here $C$ is the rational
curve $x_0=0$. The Picard group of $Z$ is rank one and the line bundle
$\O(1)$ on $Z$ restricts to $\O_C(1)$ and thus we have a line.

The other phase is an orbifold with an $A_1$ singularity, i.e.,
$\C^2/\Z_2$. Going between the phases consists of blowing up the
orbifold. Since the Picard group is rank one we may specify the
complexified K\"ahler form as a complex number $B+iJ$.  As is
well-known \cite{AGM:sd,me:enhg}, $B+iJ=\ff12$ in this orbifold
limit. The singular theory living between these phases has $B+iJ=0$.

It is also not hard to give a compact K3 example. 
Let $n=6$, $d=4$ and the pointset $\cA$ be given by coordinates
\begin{equation*}
\setlength{\arraycolsep}{3pt}
\renewcommand{\arraystretch}{0.9}
  \begin{array}{c|cccc}x_0&1&0&0&0\\x_1&1&1&0&0\\x_2&1&0&1&0\\
    x_3&1&0&0&1\\x_4&1&-6&-4&-1\\x_5&1&-3&-2&0\end{array}
\end{equation*}
Then the bigrading of $S$ (i.e., the matrix of charges) is
\begin{equation}
  Q=\begin{pmatrix}-6&3&2&0&0&1\\0&0&0&1&1&-2\end{pmatrix},
\end{equation}
with a superpotential $W=x_0F(x_1,\ldots,x_5)$, where we can choose
the Fermat
\begin{equation}
F = x_1^2+x_2^3+(x_3^{12}+x_4^{12})x_5^6. \label{eq:f1}
\end{equation}
There are four triangulations of the pointset $\cA$ and thus four
phases. These are analogous to the 4 phases of the octic threefold
\cite{CDFKM:I}. We have a large radius \CY\ phase which is an
elliptically-fibred K3 surface with a rational curve section $C$. Here
$C$ is given by $x_5=0$ and the line bundle $\O(0,1)$ restricts to
$\O_C(1)$ on $C$, which is therefore a line. 

Another phase corresponds to an orbifold given as the sextic
hypersurface in $\P^3_{\{6,4,1,1\}}$. Going from the above \CY\ phase
to this orbifold contracts $C$. Locally, near this orbifold
resolution, this example is identical to the noncompact version of the
line above.

\subsubsection{Monodromy}  \label{sss:Lmon}

In this section we review the following well-known fact:
\begin{prop}
The monodromy around the limit point corresponding to a contracted
line in a K3 surface is that of a $\Z_2$-orbifold at otherwise large
radius limit.
\end{prop}

First let us clarify what we mean by ``otherwise large radius
limit''. In the compact case we assume that the Picard number of the
K3 surface is at least 2. Let $J\in H^2(S,\R)\cap H^{1,1}(S)$ be the
K\"ahler form. The limit in question is then such that $\langle\alpha,
J\rangle\to\infty$ for any homology class $\alpha$ not proportional to the
class of our contracted curve. The noncompact case is implicitly at
this limit.

We can determine the behaviour of the phase limit induced by
contracting a line $C$ by way of the monodromy tricks that have been
employed in various examples \cite{me:TASI-D} using the elegant
formulation of spherical functors
\cite{MR2258045,Anno:sph,nick,AdAs:masscat}.  This language nicely
avoids the messy procedure of having to compute periods in form of
hypergeometric systems in more than one variable. We have a functor
\begin{equation}
\mathbf{F}:\mathbf{A} \to\DC(S),
\end{equation}
from some triangulated category $\mathbf{A}$. If this satisfies certain
conditions we may construct an automorphism $T_{\mathbf{F}}$ of
$\DC(S)$. If $\mathbf{F}$ is injective on objects we will
denote this as $T_{\mathbf{A}}$. 

To simplify notation, let us assume the Picard number is
exactly 2.  Given that we have a contractible {\em line}, there must
be two distinct primitive divisor classes, $H$ and $L$, in the K3 surface $S$
such that we have intersection numbers
\begin{equation}
  H.C=0, \qquad L.C = 1.
\end{equation}
$C.C=-2$ implies that the divisor class of $C$ is $hH-2L$
for some integer $h$.

As in \cite{AdAs:masscat}, we choose a 2-sphere in the moduli space of
complexified K\"ahler forms. This 2-sphere is near the large radius
limit for everything other than $C$. We have 3 interesting points on
the sphere:
\begin{itemize}
\item $0$: The large $C$ limit.
\item $\infty$: The limit point we go to upon contracting $C$.
\item $\Delta$: The point in the wall dividing the phases where we
have a necessarily have a singular conformal field theory. 
\end{itemize}
Let $T_0$,
$T_\infty$ and $T_\Delta$ denote the automorphisms of the derived
category induced by monodromy around these points. The topology of a
2-sphere with 3 punctures implies $T_\infty=T_\Delta T_0$.

We may choose $T_0=-\otimes\O(L)$.\footnote{The ambiguity in this
  choice was discussed in \cite{AdAs:masscat}.} We denote this
autoequivalence as $\O(L)$ for brevity.  Contracting $C$ is an
EZ-transformation of Horja \cite{Horj:EZ}, or more specifically a
Seidel--Thomas twist \cite{ST:braid}. In spherical functor language
$T_\Delta=T_{\langle \O_C\rangle}$, where $\langle \O_C\rangle$ is the
derived category of a point and the image of this point under the
spherical functor $\mathbf{F}$ is $\O_C$. In terms of physics, this
signifies that a single stable D-brane, corresponding to $\O_C$,
becomes massless at $\Delta$ \cite{AD:Dstab,me:TASI-D} thus inducing
the singular theory.

If $\Phi$ is an autoequivalence of $\DC(S)$ then \cite{AdAs:masscat}
\begin{equation}
  \Phi T_{\mathbf{F}} = T_{\Phi\mathbf{F}}\Phi.
\end{equation}

We now compute
\begin{equation}
\begin{split}
  T_\infty^2 &= T_{\Delta}T_0T_{\Delta}T_0\\
   &= T_{\langle \O_C\rangle} \O(L) T_{\langle \O_C\rangle} \O(L) \\
   &= T_{\langle \O_C\rangle} T_{\langle \O_C(1)\rangle} \O(2L)\\
   &= T_{\langle \O_C, \O_C(1)\rangle} \O(2L), 
\end{split}
\end{equation}
where the last equality follows from a result due to Kuznetsov (see
theorem 11 in \cite{AdAs:masscat}).

Now $\langle \O_C, \O_C(1)\rangle$ is the derived category of $C$ and
so $T_{\langle \O_C, \O_C(1)\rangle}$ is $T_{i_*}$ where $i:C\to S$ is
the inclusion map for a divisor. As in example 10 of
\cite{AdAs:masscat}, $T_{i_*}$ is therefore tensoring by $\O(C)$. But
$\O(C)=\O(hH-2L)$ so we end up with
\begin{equation}
  T_\infty^2 = \O(hH).
\end{equation}

Now the monodromy $\O(hH)$ is what we would expect for a large
radius limit. Note that $H$ does not intersect $C$ and so this is some
``large radius'' for ``the rest'' of the K3 surface away from the line
$C$. The fact we need to go {\em twice\/} around $\infty$ corresponds
to a quantum $\Z_2$-symmetry that we would expect from an
orbifold. Thus we have proved the proposition.

In K\"ahler form language, the Poincar\'e dual of the complexified
K\"ahler form can be written as
\begin{equation}
  B+iJ = xH + yL,
\end{equation}
were we go to the ``otherwise large radius limit'' by setting $x\to\infty$.
As shown in \cite{AGM:sd}, the point $\Delta$ corresponds to $y=0$; and
$\infty$ corresponds to $y=\ff12$. That is, the $B$-field on $C$ is zero for
the singular conformal field theory and $\ff12$ for the orbifold.

Note that the 2-sphere we chose in the moduli space with the points
$0$, $\infty$ and $\Delta$ needs to be deformed so that it generically
lies in the interior of the moduli space as discussed in
\cite{AdAs:masscat}. This process involves a choice that varies $h$
and so $h$ is not uniquely defined.  Note also that in the noncompact
example of the total space of $\O_C(-2)$ we effectively ignore $H$ and
get $T_\infty^2 = 1$.

\subsection{A Conic Example}   \label{ss:conic}    

Let $n=6$, $d=4$ and the pointset $\cA$ be given by
\begin{equation*}
\setlength{\arraycolsep}{3pt}
\renewcommand{\arraystretch}{0.9}
  \begin{array}{c|cccc}x_0&1&0&0&0\\x_1&1&-1&0&0\\x_2&1&1&0&0\\x_3&1&0&1&0\\
    x_4&1&0&0&1\\x_5&1&-1&-1&-1\end{array}.
\end{equation*}
Then the matrix of charges is
\begin{equation}
  Q=\begin{pmatrix}-2&-1&0&1&1&1\\-2&1&1&0&0&0\end{pmatrix},
             \label{eq:Q6}
\end{equation}
with a superpotential $W=x_0F(x_1,\ldots,x_5)$, where 
\begin{equation}
F = x_1^2f_4(x_3,x_4,x_5)+x_1x_2f_3(x_3,x_4,x_5)+
x_2^2f_2(x_3,x_4,x_5), \label{eq:f2}
\end{equation}
and $f_k$ is a homogeneous equation of degree $k$. Let us assume
that these $f_k$ are generic to avoid unnecessary singularities.

There are four triangulations of the pointset $\cA$ and thus four
phases. Two of these phases are ``hybrid'' models which can be viewed
as a fibration over a weighted projective space with fibre given by a
Landau--Ginzburg theory. We will not be concerned with these hybrid
phases in this paper.

Another phase is a large radius ``\CY'' phase where
$B=(x_1,x_2)(x_3,x_4,x_5)$. Here the ambient toric variety $Z$ is a
$\P^1$-bundle over $\P^2$. More precisely it is the bundle
$\P(\O\oplus\O(-1))$ over $\P^2$. The constraints $\partial W/\partial
x_i=0$ force $x_0=F=0$. This yields that $X$ is a double cover over $\P^2$
branched over the sextic $f_2f_4-f_3^2=0$. This is a smooth K3 surface
and so we will call this the ``smooth phase''.

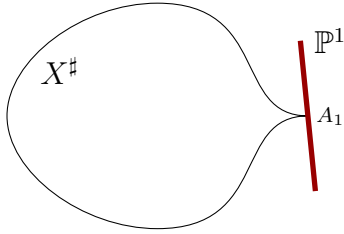
\begin{figure}
\begin{center}
\begin{tikzpicture}[scale=1.0]
\draw[yscale=0.75,xshift=5.5cm] (-5,0) .. controls (-5,1) and  (-4,2) .. (-3,2) ..
    controls (-1.5,2) and (-2,0) .. (-1,0) ..
    controls (-2,-0) and (-1.5,-2) .. (-3,-2) .. 
    controls (-4,-2) and (-5,-1) .. (-5,0);
\draw (1.2,0.6) node {$X^\sharp$};
\draw[xshift=5.5cm,color=red!60!black,line width=2pt] (-1.1,1) --
(-0.9,-1);
\draw (4.8,1) node {$\P^1$};
\draw (4.8,0) node {$\scriptstyle A_1$};
\end{tikzpicture}
\end{center}
\caption{An exoflop.}   \label{fig:exoflop}
\end{figure}

The other phase is the ``exoflop'' phase and has
\begin{equation}
  B = (x_0,x_1)\cap(x_1,x_2)\cap(x_2,x_3,x_4,x_5).  \label{eq:Bexo}
\end{equation}
The critical point set of $W$ in this toric variety is reducible and
has two components. One component has $x_0=0$. From (\ref{eq:Bexo}) this
forces $x_1\neq0$ and we may use one $\C^*$-action to set
$x_1=1$. This leaves a quartic K3 surface in $\P^3$
with homogeneous coordinates $[x_2,x_3,x_4,x_5]$ given by
\begin{equation}
f_4(x_3,x_4,x_5)+x_2f_3(x_3,x_4,x_5)+x_2^2f_2(x_3,x_4,x_5)=0.
\end{equation}
This has an $A_1$ singularity at $[1,0,0,0]$. The other component is a
weighted projective space $\P^1_{\{2,1\}}$, which is isomorphic to
$\P^1$, with homogeneous coordinates $[x_0,x_1]$ and with
$[x_2,x_3,x_4,x_5]=[1,0,0,0]$. The two components thus intersect at
the $A_1$ point. The critical point set of $W$ looks like
figure~\ref{fig:exoflop}, where $X^\sharp$ is the singular quartic K3.

One views the transition from the smooth phase to the exoflop phase as
shrinking down the rational curve $C\subset X$ given by
$x_1=0$. Then, after it has shrunk to point, one continues into the
exoflop phase by ``growing'' the external $\P^1$ out of the resulting
singularity. 

The curve $C$ is given by $f_2(x_3,x_4,x_5)=0$ and is manifestly a
conic in $\P^2$. Furthermore $\O(a,b)$ restricts to $\O_C(2a)$ so $C$
is a conic by definition \ref{def:conic} too.

\subsubsection{Monodromy} \label{sss:cmon}

In this section we prove the following general result:
\begin{prop}
  The monodromy around the limit point corresponding to a contracted
  conic in a K3 surface is the same as the monodromy around a singular
  conformal field theory equivalent to that of contracted line (with
  zero $B$-field) at otherwise large radius limit.
\end{prop}

Again we can use the language of spherical functors to calculate
monodromies. Now we have divisor classes $H$ and $L$ such that
\begin{equation}
  H.C=0, \qquad L.C = 2.
\end{equation}
Furthermore, $C.C=-2$ implies that the divisor class of $C$ is $hH-L$
for some integer $h$.

Then
\begin{equation}
\begin{split}
  T_\infty &= T_\Delta T_0\\
          &= T_{\langle\O_C\rangle}\O(L)\\
          &= T_{\langle\O_C\rangle}\O(C)^{-1}\O(hH)\\
          &=
          T_{\langle\O_C\rangle}T_{\langle\O_C(-1),\O_C\rangle}^{-1}\O(hH)\\
          &= T_{\langle\O_C\rangle}T_{\langle\O_C\rangle}^{-1}
          T_{\langle\O_C(-1)\rangle}^{-1}\O(hH)\\
          &= T_{\langle\O_C(-1)\rangle}^{-1}\O(hH)
\end{split}
\end{equation}
Again, as in section \ref{sss:Lmon} we have a large radius monodromy
$\O(hH)$ coming from geometry away from the contraction. The more
interesting part of the monodromy is the Seidel--Thomas
\cite{ST:braid} twist $T_{\langle\O_C(-1)\rangle}$, which is the
monodromy associated to a single object $\O_C(-1)$ becoming massless.
(Note the direction of the monodromy around $\infty$ is opposite to
that around $\Delta$ and thus we have the inverse of a Seidel--Thomas
twist for a massless D-brane.)

The nature of monodromy around $\infty$ is therefore identical (except for
global geometric issues) the to monodromy around $\Delta$. That is, we
have a single massless D-brane, $\O_C$ at the discriminant $\Delta$
and another single massless D-brane $\O_C(-1)$ at the exoflop limit
point $\infty$. 

The $B$-field language is fairly clear. In the case of the line, $B=0$
is at $\Delta$ and $B=\ff12$ is at $\infty$. In the case of the conic
we effectively double the $B$-field and so $B=0$ at the $\Delta$ but
now $B=1$ at $\infty$. But $B=1$ is the equivalent singular conformal
field theory to $B=0$ except that we need to relabel $D$-branes by a
shift $-\otimes\O(-1)$. This is exactly what we see above. The
location of the orbifold point at $B=\ff12$ is not manifestly obvious in
the exoflop picture.


\section{Geometrical Picture of D-branes on an Exoflop}
    \label{s:geom}

We see that the contracted conic curve produces a massless D-brane
$\O_C(-1)$ in the exoflop limit. But denoting it $\O_C(-1)$ is using
the language of the large radius smooth phase. What does this massless
D-brane look like in terms of the exoflop phase? Rather than prove
general statements, we will analyze the example of section
\ref{ss:conic} and then a more complicated example in section
\ref{ss:eg2}. We assume other cases will be similar.

\subsection{The conic}  \label{ss:conicc}

First let us review the basic algebraic geometry of a conic. Consider
$C\subset\P^2$ given by $f=z_0^2-z_1z_2=0$. A coherent sheaf $\tilde
M$ on $C$ is associated to an $R$-module, $M$, where
\begin{equation}
  R = \frac{S}{(f)}, \quad S=\C[z_0,z_1,z_2].
\end{equation}
Clearly the module $R(n)$ yields $\O_C(2n)$. So which modules
give $\O_C(\textrm{odd})$? The skyscraper sheaf of the point $[0,0,1]$
is given by the module $\coker(z_0,z_1)$. This has an infinite free resolution
\begin{equation}
\xymatrix@C=14mm{
  \cdots\ar[r]^-{\left(\begin{smallmatrix}z_0&z_1\\z_2&z_0
               \end{smallmatrix}\right)}&
  R(-4)^{\oplus2}\ar[r]^-{\left(\begin{smallmatrix}z_0&-z_1\\-z_2&z_0
               \end{smallmatrix}\right)}&
  R(-3)^{\oplus2}\ar[r]^-{\left(\begin{smallmatrix}z_0&z_1\\z_2&z_0
               \end{smallmatrix}\right)}&\\
  &R(-2)^{\oplus2}\ar[r]^-{\left(\begin{smallmatrix}z_0&-z_1\\-z_2&z_0
               \end{smallmatrix}\right)}&
  R(-1)^{\oplus2}\ar[r]^-{\left(\begin{smallmatrix}z_0&z_1
               \end{smallmatrix}\right)}&
  R\ar[r]&M\ar[r]&0.}
\end{equation}
But we also have
\begin{equation}
\xymatrix@1{
  0\ar[r]&\O_C(-1)\ar[r]&\O_C\ar[r]&\O_{\textrm{\tiny pt}}\ar[r]&0,}
\end{equation}
and so (mixing sheaves and modules)
\begin{equation}
\xymatrix@1@C=15mm{
  \cdots\ar[r] 
   &R(-2)^{\oplus2}\ar[r]^-{\left(\begin{smallmatrix}z_0&-z_1\\-z_2&z_0
               \end{smallmatrix}\right)}&
   R(-1)^{\oplus2}\ar[r]&\O_C(-1)\ar[r]&0.}
\end{equation}
We may tensor this resolution by $R(n)$ to resolve $\O_C(2n-1)$. In
the usual way \cite{Eis:mf} we associate infinite resolutions with
matrix factorizations. This
means that $\O_C(\textrm{odd})$ is associated to the graded matrix factorization
\begin{equation}
\begin{pmatrix}z_0&z_1\\z_2&z_0\end{pmatrix}
\begin{pmatrix}z_0&-z_1\\-z_2&z_0\end{pmatrix}
= f.\id,  \label{eq:MF1}
\end{equation}
of suitable degree.

\subsection{Following $\O_C(-1)$}

We can now use the above construction to build a matrix factorization
for $\O_C(-1)$ in the K3 surface $S$. We can then use the well-known
procedure of carrying this over to the exoflop phase and reinterpret it
there. That is, we have an equivalence of D-brane categories from the
smooth K3 phase to the exoflop phase:
\def\DCexo{\mathbf{D}_{\mathrm{exo}}}
\begin{equation}
\Phi:\DC(S) \to \DCexo.
\end{equation}
We would like to compute $\Phi(\O_C(-1))$.

The procedure of moving objects in the derived category between
different phases has been discussed in
\cite{Orlov:mfc,HHP:linphase,me:csalg,MR2795327}. Let
$S=\C[x_0,x_1,\ldots]$ be the homogeneous coordinate ring of the toric
variety $Z$. Assume we have a
superpotential $W=x_0F(x_1,x_2,\ldots)$ and let $X$ be the \CY\
manifold $x_0=F=0$.

Beginning in the \CY\ phase we have a coherent sheaf which is
described as an $A$-module $M$, where
\begin{equation}
  A = \frac{\C[x_1,x_2,\ldots]}{(F)}.
\end{equation}
This $A$-module can be associated with a graded matrix
factorization. The procedure for doing this is a little tedious but is
easily computed via computer algebra as explained in detail in
\cite{me:csalg}. Indeed, the procedure is built into Macaulay 2 as
part of the mechanism for computing Ext groups \cite{AG:McExt}. The
resulting matrix factorization consists of maps between free graded
$S$-modules enhanced by an extra grading. This new degree is
associated with R-symmetry of the underlying conformal field theory or
is considered a ``homological'' grading. This process of producing a
matrix factorization also reintroduces the variable $x_0$ (called
$X_1$ in \cite{AG:McExt}). The triply-graded degrees of the variables
are now
\begin{equation}
\begin{array}{c|ccc}
&R&Q_1&Q_2\\
\hline
x_0&2&-2&-2\\
x_1&0&-1&1\\
x_2&0&0&1\\
x_3&0&1&0\\
x_4&0&1&0\\
x_5&0&1&0
\end{array}  \label{eq:3deg}
\end{equation}
We can rephrase this data by taking the summand of the
two free modules and writing a single matrix acting as an endomorphism
on this larger free module. Thus we write a graded matrix
factorization as
\begin{equation}
  u : \bigoplus_{i=1}^{2s} S(\mathbf{q}_i) \to
    \bigoplus_{i=1}^{2s} S(\mathbf{q}_i).   \label{eq:mfu}
\end{equation}
The matrix representing $u$ squares to $W$. The two original
rank $s$ $S$-modules may be reconstructed by taking even and odd homological
degrees of the big module.  Note that any finite complex of sheaves can
similarly be converted into a finite matrix factorization.

The next ingredient is to find a tilting collection
\begin{equation}
  T = \bigoplus_{i=1}^r S(\mathbf{t}_i),
\end{equation}
where the degrees $\mathbf{t}_i$ fit inside an $r$-dimensional
``window'' to pass between the phases as described in
\cite{HHP:linphase}. Any matrix factorization of the form
(\ref{eq:mfu}) is unchanged as it passes between the phases if all the
degrees $\mathbf{q}_i$ are elements of $\{\mathbf{t}_1, \mathbf{t}_2,
\ldots, \mathbf{t}_r\}$.

More generally we can manipulate the matrix factorization
(\ref{eq:mfu}) to make an object whose degrees $\mathbf{q}_i$ fit inside the
window. We can doing this by taking mapping cones to and from objects
which are trivial thanks to the irrelevant ideal. This construction
describes how to transport general objects between phases. 

For definiteness let us use a specific smooth
\begin{equation}
  F = x_1^2(x_3^4+x_4^4+x_5^4) + x_2^2(x_3x_4-x_5^2).
\end{equation}
for the example from section \ref{ss:conic}. We will
prove the following key result
\begin{prop}
The matrix factorization obtained from $\O_C(-1)$ in the
smooth phase passes unchanged into the exoflop phase. That is, the
same matrix factorization that Macaulay 2 gives us can be used for
this object in $\DCexo$.
\end{prop}

The tilting collection we
want to use to be consistent with section \ref{sss:cmon} is given by%
\footnote{By consistent we mean that the monodromy $T_\Delta$ is
  correctly determined by the tilting collection. The process of
  deriving $T_\Delta$ from a tilting collection was described in
  \cite{HHP:linphase} (see also \cite{shipping}). One essentially
  computes $T_\Delta=T_0^{-1}T_\infty$ where both $T_0$ and $T_\infty$
  are given by the same degree shift. For this case we use a degree
  shift of $(1,-1)$.}
\begin{equation}
\xymatrix@R=0mm@C=0mm{
S(-2,-1)&S(-1,-1)&S(0,-1)\\
S(-2,-2)&S(-1,-2)&S(0,-2)}
\end{equation}

Here we have the conic $C$ given by $x_1=0$. Thus $\O_C$ is (the sheaf
associated to the module given as) the cokernel of the map
$(x_1)$. The cokernel of $(x_1\; x_3\; x_5)$ gives the point
$x_1=x_3=x_5$ on $C$. Therefore, as in section \ref{ss:conicc}, the
module associated to $\O_C(-1)$ is the kernel of the quotient map
\begin{equation}
k:\coker(x_1)\to\coker(x_1\; x_3\; x_5).
  \label{eq:kck}
\end{equation}
Feeding this module into Macaulay gives the matrix factorization:

\begin{samepage}
\begin{Verbatim}[fontsize=\scriptsize]
{0, -1, 0}   | 0         0        -x_0x_1x_4^4-x_0x_1x_5^4 -x_0x_1x_3^3x_5                      
{0, -1, 0}   | 0         0        0                        -x_0x_1x_3^4-x_0x_1x_4^4-x_0x_1x_5^4 
{-1, 0, -1}  | x_1       0        0                        0                                    
{-1, 0, -1}  | 0         x_1      0                        0                                     ...
{-1, -2, 0}  | x_3       -x_5     0                        0                                    
{-1, -2, -2} | -x_2^2x_5 x_2^2x_4 0                        0                                    
{-2, -1, -1} | 0         0        -x_3                     x_5                                  
{-2, -1, -3} | 0         0        x_2^2x_5                 -x_2^2x_4                            

\end{Verbatim}
\vspace{-11mm}
\hspace{9mm}$\underbrace{\phantom{,,,}}$\\
$\phantom{xxxxx}${\scriptsize $\mathbf{q}_i$}
\end{samepage}

\smallskip

We do not care about the contents of this matrix, only the degrees.
The degrees $\mathbf{q}_i$ are read off as the second and third
entries as shown above (the first being the homological degree).
Clearly they do not all lie in the tilting collection so we need to
manipulate this object to fit it into the window.

The irrelevant ideal of both the \CY\ phase and the exoflop phase is
contained in the ideal $(x_1,x_2)$. It follows that the complex
\begin{equation}
\xymatrix@1@C=11mm{
  0\ar[r]&
    S(1,-2)\ar[r]^-{\left(\begin{smallmatrix}-x_2\\x_1
          \end{smallmatrix}\right)}
      &{\begin{array}{c}S(1,-1)\\\oplus\\S(0,-1)\end{array}}
    \ar[r]^-{\left(\begin{smallmatrix}x_1&x_2\end{smallmatrix}\right)}
           &S\ar[r]&0,}  \label{eq:x1x2}
\end{equation}
is annihilated by the irrelevant ideal in both phases and is therefore
trivial in the D-brane category in both phases. We now use this
complex, and its degree shifts, to extend our window. That is, if three of
the four $S(\mathbf{q}_i)$'s in (\ref{eq:x1x2}) (or its shifts) appear in
our tilting collection then we will add the fourth one. Then iterate this
process.

\begin{figure}
\begin{center}
\begin{tikzpicture}[scale=0.5]
\draw[step=1cm,gray,very thin] (-3.9,-4.9) grid (2.9,1.9);
\filldraw (0,-1) circle [radius=0.1];
\filldraw (-1,-1) circle [radius=0.1];
\filldraw (-2,-1) circle [radius=0.1];
\filldraw (0,-2) circle [radius=0.1];
\filldraw (-1,-2) circle [radius=0.1];
\filldraw (-2,-2) circle [radius=0.1];
\filldraw (-2,0) circle [radius=0.1];
\filldraw (-1,0) circle [radius=0.1];
\filldraw (-2,1) circle [radius=0.1];
\filldraw (0,-3) circle [radius=0.1];
\filldraw (-1,-3) circle [radius=0.1];
\filldraw (0,-4) circle [radius=0.1];
\draw (0,0) circle [radius=0.2];
\end{tikzpicture}
\end{center}
\caption{Extended window for the exoflop transition.} \label{fig:ewindow}
\end{figure}
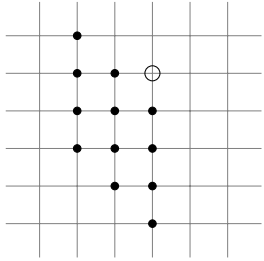

This procedure leads to a collection of $\mathbf{q}_i$'s as shown in
figure \ref{fig:ewindow}, where the circle represents the origin. 

Now all the degrees appearing in the matrix factorization for
$\O_C(-1)$ appear in this collection. This means we can build the
matrix factorization out of objects in the tilting collection and
objects which are trivial in both phases. This means we can copy this
matrix factorization unchanged from the smooth phase into the
exoflop phase and we prove the proposition.

\subsection{Locating $\O_C(-1)$}

Since the matrix factorization for $\O_C(-1)$ passed unchanged to the
exoflop phase we can retain its interpretation as the kernel of the
map (\ref{eq:kck}). This allows us to locate it precisely. In
particular it is supported at $x_1=0$. The irrelevant ideal
(\ref{eq:Bexo}) forces $x_0\neq0$. Since the K3 surface component of
the exoflop has $x_0=0$ we see that this D-brane is not located there.

The other component of the exoflop, the $\P^1$ sticking out, has
homogeneous coordinates $[x_0,x_1]$ and thus the D-brane $\O_C(-1)$ is
located on here at the ``other end'' of the $\P^1$ from where it
attached to the K3 component. We depict this in figure \ref{fig:OCfound}.
From now on we will use ``North pole'' to denote the location of
$\O(-1)$ on $\P^1$ and ``South pole'' to denote the intersection with
$X^\sharp$. 

\begin{figure}
\begin{center}
\begin{tikzpicture}[scale=1.0]
\draw[yscale=0.75,xshift=5.5cm] (-5,0) .. controls (-5,1) and  (-4,2) .. (-3,2) ..
    controls (-1.5,2) and (-2,0) .. (-1,0) ..
    controls (-2,-0) and (-1.5,-2) .. (-3,-2) .. 
    controls (-4,-2) and (-5,-1) .. (-5,0);
\draw (1.2,0.6) node {$X^\sharp$};
\draw[xshift=5.5cm,color=red!60!black,line width=2pt] (-1.1,1.5) --
(-0.9,-1);
\draw (5.0,-1) node {$\P^1$};
\filldraw (4.42,1.3) circle [radius=0.07];
\draw (6.5,1) node {$\O_C(-1)$};
\draw[->] (5.7,1.1) -- (4.55, 1.3);
\draw (6.3,0) node {$\O_C$};
\draw[->] (6.0,0.1) -- (4.55, 1.1);
\draw[->] (6.0,0.1) -- (4.59, 0.5);
\draw[->] (6.0,0.1) -- (4.64, -0.5);
\end{tikzpicture}
\end{center}
\caption{$\O_C(-1)$ located.}   \label{fig:OCfound}
\end{figure}

Another case of interest corresponds to the cokernel of
$(x_0\,x_1)$. This module corresponds to $\O_C$ in the smooth phase. A
resolution of this module does not fit in the window or even the
extended window of figure \ref{fig:ewindow}. We are forced to
``prepare it'' by applying a mapping cone with $\coker(x_3\,x_4\,x_5)$
to get it into the window. Now when we pass into the exoflop phase,
the object $\coker(x_3\,x_4\,x_5)$ is supported all along the external
$\P^1$. Thus, $\O_C$ is not supported just at the North pole --- it is
spread over the whole $\P^1$ as depicted in figure \ref{fig:ewindow}.

It is worth pointing out that no object can be localized at a single
point on the external $\P^1$ other than the poles. The homogeneous
coordinates are $[x_0,x_1]$. It follows that any such object would be
associated with a module as the cokernel of a matrix involving
$ax_0+bx_1$, where $a$ and $b$ are nonzero and might involve other
coordinates. Such any expression can never be homogeneous with respect
to the grading (\ref{eq:3deg}).  We therefore see that the exoflop is
unphysical in the sense that no point-like D-brane can ``move'' along
the external $\P^1$.

\subsection{The \GLSM\ paradigm}   \label{ss:glsm}

We have argued that, in the case of contracting a conic, the
singular conformal field theory appearing in the wall between the
phases is essentially identical to that of the phase limit. Both
theories involve a single D-brane becoming massless. Singularities
appear in the gauged linear $\sigma$-model because the space on which
the theory is defined becomes noncompact.

The potential for bosonic fields is \cite{W:phase,MP:inst}
\begin{equation}
U = \sum_a\frac{D_a}{2e^2} + 2\sum_{a,b}\bar\sigma_a\sigma_b
  \sum_iQ^a_iQ^b_i|x_i|^2 +
  \sum_i\left|\frac{\partial W}{\partial x_i}\right|^2.
\end{equation}
There are two standard ways the vacuum for $U=0$ can become noncompact
and then we also have the exoflop way as we explain below.

The first type of singularity is the prototypical way we expect
singularities to appear in the K\"ahler moduli space.
As we move in the moduli space of complexified K\"ahler forms we vary
parameters in $D_a$. Setting $D_a=0$ then fixes values for
$|x_i|^2$. At the walls in the phase picture we have $|x_i|^2=0$ and
thus some of the $\sigma$ fields become unconstrained and
noncompact. This is why we have singularities along the discriminant
$\Delta$ as explained in detail in \cite{MP:inst}.

The second type of singularity is the prototypical way we expect
singularities to appear in the complex structure moduli space. In this
case we decompactify the vacuum by deforming the complex structure to
acquire a singularity.  If $W=x_0F(x_1,x_2,\ldots)$ then
\begin{equation}
\sum_{i=0}\left|\frac{\partial W}{\partial x_i}\right|^2
=
|F|^2 + |x_0|^2\sum_{i=1}\left|\frac{\partial F}{\partial x_i}\right|^2.
\end{equation}
Thus, at a singularity where $F=\partial F/\partial x_i=0$ the field
associated to $x_0$ can go off to infinity.

The singularity at the exoflop limit is essentially equivalent to this
latter complex-structure-induced singularity but it is obtained by
varying the K\"ahler form. In the exoflop phase, the $X^\sharp$
component is singular, i.e., it has a solution of $F=\partial
F/\partial x_i=0$. Thus $x_0$ is liberated here which is why we have
another $\P^1$ component attached at this point. But this $\P^1$ is
{\em compact\/} and so the theory is not singular until we go all the
way to the exoflop limit. In this limit the area of the external
$\P^1$ is infinite (in the \GLSM\ picture) and so the vacuum is not
compact. 

\subsection{The noncommutative resolution paradigm} \label{ss:ncr}

The noncommutative resolution of \cite{bergh04:nc} is a purely
algebraic way of resolving a singularity. The idea is as follows (see
also \cite{Aspinwall:2010mw}). Suppose the singular variety is affine
$X=\Spec R$. Then the fact that $X$ is singular is reflected in the
fact that $R$ has infinite global dimension. That is, there is an
$R$-module $M$ that does not have a finite free resolution. We then
use $M$ to ``blow-up'' $X$.

For simplicity assume $X$ is a hypersurface singularity. (Indeed
these are the only examples studied to date.) Then the
infinite free resolution of $M$ corresponds to a matrix
factorization. We can then define and analyze the algebra
\begin{equation}
  A = \End(R\oplus M).
\end{equation}
If this algebra has finite global dimension we are done. If not, we
repeat the process and add in more modules.

When the process is complete we expect an equivalence of categories
\begin{equation}
\DC(\tilde X) \simeq \DC(\textrm{mod-}A),
\end{equation}
where $\tilde X$ is a crepant resolution of $X$. Note that $R\oplus M$
is acting like a tilting object in this case, and that tilting objects
cannot exist on a compact \CY\ because of Serre duality. Thus
noncommutative resolutions would seem necessarily restricted to
describing noncompact geometries.

In the case of an $A_1$ singularity $z_0^2-z_1z_2=0$ in $\C^3$ the required
matrix factorization for $M$ is (\ref{eq:MF1}).

We claim the exoflop gives a very satisfying geometric description of
the this non-geometric construction and indicates how to fit
noncommutative resolutions into compact \CY's. In our K3 case the
matrix factorization that ``smooths'' the singularity in $X^\sharp$ is
precisely the D-brane living at the North pole of the $\P^1$.


\subsection{Exoflop Chains}  \label{ss:eg2}

We consider another example to get a better idea of the generic
appearance of exoflops in K3 surfaces.
Let $n=7$, $d=4$ and the pointset $\cA$ be given by coordinates
\begin{equation*}
\setlength{\arraycolsep}{3pt}
\renewcommand{\arraystretch}{0.9}
  \begin{array}{c|cccc}x_0&1&0&0&0\\x_1&1&-2&-1&0\\x_2&1&-1&-1&-1\\
    x_3&1&-1&0&1\\x_4&1&0&0&1\\x_5&1&0&1&0\\x_6&1&1&0&0\end{array}
\end{equation*}
Then the trigrading of $S$ is
\begin{equation}
  Q=\begin{pmatrix}-2&0&0&1&0&0&1\\0&1&0&-2&0&1&0
     \\-2&-1&1&1&1&0&0\end{pmatrix},
\end{equation}
with a superpotential $W=x_0F(x_1,\ldots,x_5)$. If we consider only
the monomials at the vertices of the Newton polytope this polynomial
would be
\begin{equation}
F = x_1^4x_2^4x_3^2 + x_1^4x_3^2x_4^4 + x_3^2x_5^4
  + x_2^2x_6^2 + x_4^2x_6^2. \label{eq:f3}
\end{equation}
 
The K\"ahler moduli space has dimension $r=3$. The pointset has 12
triangulations. One of these phases yields $X_\Sigma$ as a K3 surface
given as a double cover of a Hirzebruch surface ${\mathbb F}_1$
branched over a suitable curve. 

\begin{figure}
\begin{center}
\begin{tikzpicture}[x=0.22mm,y=0.22mm]
 \path[shape=circle,inner sep=1pt,every node/.style={draw}]
     (253,236) node(a1) {$\phantom{1}$}
     (188,245) node(a2) {$\phantom{1}$}
     (495,266) node(a3) {$\phantom{1}$}
     (615,301) node(a4) {$\phantom{1}$}
     (447,128) node(a5) {5}
     (394,301) node(a6) {$\phantom{1}$}
     (575,177) node(a7) {4}
     (359,56) node(a8) {3}
     (344,16) node(a9) {1}
     (298,42) node(a10) {$\phantom{1}$}
     (469,69) node(a11) {2}
     (415,90) node(a12) {$\phantom{1}$};
  \draw (a1) -- (a2) -- (a6) -- (a4) -- (a3) -- (a1)
     -- (a8) -- (a5) -- (a7) -- (a11) -- (a9) -- (a8);
  \draw (a10) -- (a12) -- (a11) -- (a9) -- (a10) -- (a2);
  \draw (a4) -- (a7);
  \draw (a12) -- (a6);
  \draw (a5) -- (a3);
\end{tikzpicture}
\end{center}
\caption{The secondary polytope for section \ref{ss:eg2}.}  \label{fig:eg2}
\end{figure}
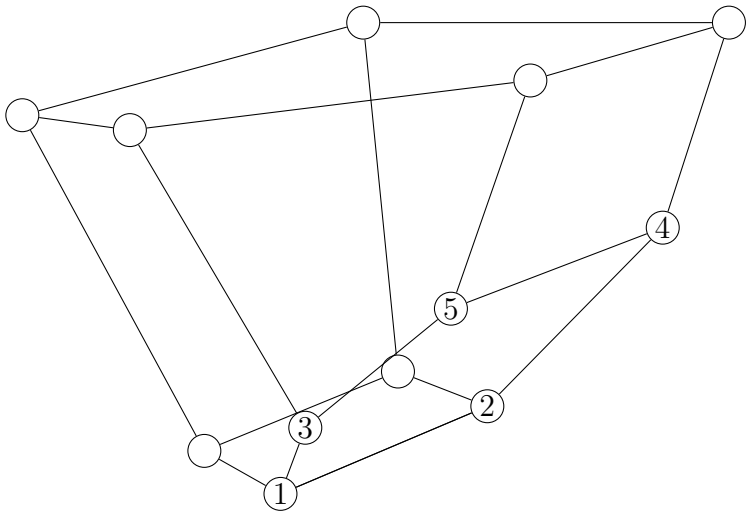

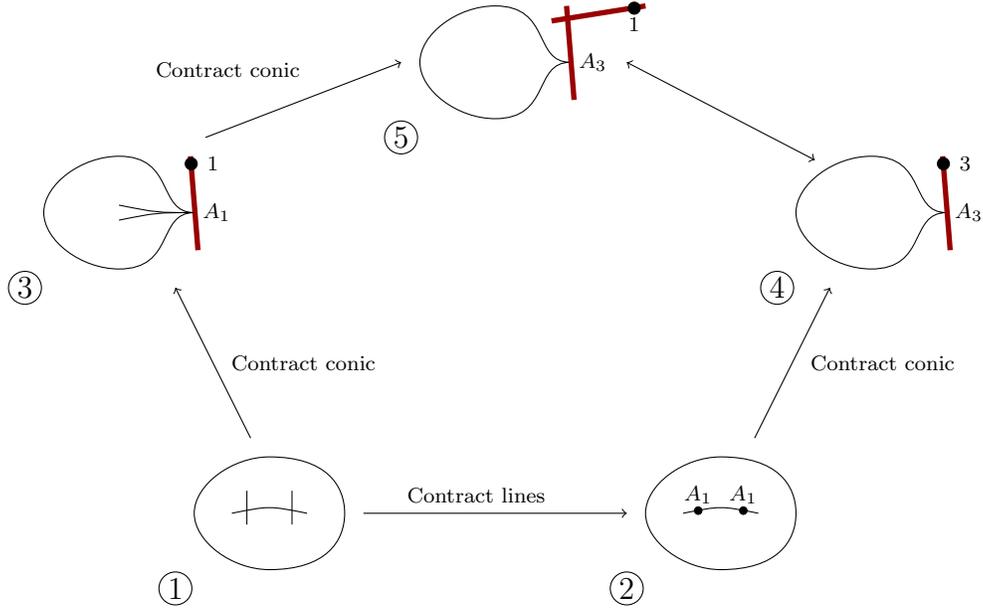
\begin{figure}
\begin{center}
\begin{tikzpicture}[scale=1.0]
\def\exoflopg{
\draw[yscale=0.75,xshift=5.5cm] (-5,0) .. controls (-5,1) and  (-4,2) .. (-3,2) ..
    controls (-1.5,2) and (-2,0) .. (-1,0) ..
    controls (-2,-0) and (-1.5,-2) .. (-3,-2) .. 
    controls (-4,-2) and (-5,-1) .. (-5,0);
\draw[xshift=5.5cm,color=red!60!black,line width=2pt] (-1.1,1.5) --
(-0.9,-1);}
\begin{scope}[xshift=-50mm,yshift=20mm,scale=0.5]
\exoflopg
\filldraw (4.42,1.3) circle [radius=0.16];
\draw(5,1.3) node {\scriptsize 1};
\draw(4.5,0) .. controls (3.5,0) .. (2.5,0.2); 
\draw(4.5,0) .. controls (3.5,0) .. (2.5,-0.2); 
\draw(5.1,0) node {$\scriptstyle A_1$};
\path[shape=circle,inner sep=1pt,every node/.style={draw}]
  (0,-2) node {3};
\end{scope}
\begin{scope}[xshift=0mm,yshift=40mm,scale=0.5]
\exoflopg
\draw[xshift=5.5cm,color=red!60!black,line width=2pt] (-1.5,1.1) --
(1,1.5);
\draw(5.1,0) node {$\scriptstyle A_3$};
\filldraw (6.2,1.44) circle [radius=0.16];
\draw(6.2,1) node {\scriptsize 1};
\path[shape=circle,inner sep=1pt,every node/.style={draw}]
  (0,-2) node {5};
\end{scope}
\begin{scope}[xshift=50mm,yshift=20mm,scale=0.5]
\exoflopg
\filldraw (4.42,1.3) circle [radius=0.16];
\draw(5,1.3) node {\scriptsize 3};
\draw(5.1,0) node {$\scriptstyle A_3$};
\path[shape=circle,inner sep=1pt,every node/.style={draw}]
  (0,-2) node {4};
\end{scope}
\def\smoothgg{
\draw[yscale=0.75,xshift=5.5cm] (-5,0) .. controls (-5,1) and  (-4,2) .. (-3,2) ..
    controls (-1.5,2) and (-1,1) .. (-1,0) ..
    controls (-1,-1) and (-1.5,-2) .. (-3,-2) .. 
    controls (-4,-2) and (-5,-1) .. (-5,0);
}
\begin{scope}[xshift=30mm,yshift=-20mm,scale=0.5]
\smoothgg
\draw(1.5,0) .. controls (2.5,0.2) .. (3.5,0);
\draw (1.9,0.5) node {$\scriptstyle A_1$};
\filldraw(1.9,0.07) circle [radius=0.1];
\draw (3.1,0.5) node {$\scriptstyle A_1$};
\filldraw(3.1,0.07) circle [radius=0.1];
\path[shape=circle,inner sep=1pt,every node/.style={draw}]
  (0,-2) node {2};
\end{scope}
\begin{scope}[xshift=-30mm,yshift=-20mm,scale=0.5]
\smoothgg
\draw(1.5,0) .. controls (2.5,0.2) .. (3.5,0);
\draw(1.9,-0.3) -- (1.9,0.6);
\draw(3.1,-0.3) -- (3.1,0.6);
\path[shape=circle,inner sep=1pt,every node/.style={draw}]
  (0,-2) node {1};
\end{scope}
\draw[->] (-2,-1) -- (-3,1);
\draw (-2.4,0) node [anchor=west] {\scriptsize Contract conic};
\draw[->] (-2.6,3) -- (0,4);
\draw (-1.2,3.9) node [anchor=east] {\scriptsize Contract conic};
\draw[->] (-0.5,-2) -- (3,-2);
\draw (1,-2) node [anchor=south] {\scriptsize Contract lines};
\draw[->] (4.7,-1) -- (5.7,1);
\draw (5.3,0) node [anchor=west] {\scriptsize Contract conic};
\draw[<->] (5.5,2.7) -- (3,4);
\end{tikzpicture}
\end{center}
\caption{A face of the secondary polytope.} \label{fig:face}
\end{figure}

The 3-dimensional secondary polytope in this case is shown in figure
\ref{fig:eg2}. It is interesting to focus on one of the pentagonal
faces of this polytope as shown in figure \ref{fig:face}. These phases
are as follows
\begin{itemize}
  \item[1:] $B=(x_1,x_5)\cap(x_1,x_6)\cap(x_3,x_6)\cap(x_2,x_3,x_4)
  \cap(x_2,x_4,x_5)$. This phase is a smooth K3 (a flop of the double
  cover of ${\mathbb F}_1)$.
  \item[2:] $B=(x_3)\cap(x_1,x_6)\cap(x_2,x_4,x_5)$. A singular K3
    with two distinct $A_1$ singularities.
  \item [3:] $B = (x_0,x_1)\cap(x_1,x_5)\cap(x_1,x_6)\cap
          (x_3,x_6)\cap(x_2,x_4,x_5)$. An exoflop. One component is a
          K3 surface which is a double cover branched over a nodal
          sextic. This surface has an $A_1$ singularity. The other
          component is isomorphic to $\P^1$ with coordinates $[x_0,x_1]$.
  \item [4:]
    $B=(x_3)\cap(x_0,x_1)\cap(x_1,x_6)\cap(x_2,x_4,x_5,x_6)$. An
    exoflop. One component is a quartic K3 surface with an $A_3$
    singularity. The other component is a $\P^1$ with homogeneous
    coordinates $[x_0,x_1]$.
  \item [5:]
    $B=(x_0,x_1)\cap(x_0,x_3)\cap(x_1,x_5)\cap(x_1,x_6)\cap(x_3,x_6)
           \cap(x_2,x_4,x_5,x_6)$. A ``double exoflop''. One component
           is a quartic K3 surface with an $A_3$ singularity. There are two
           other components each isomorphic to $\P^1$ forming a chain
           off this singular point. These $\P^1$'s have coordinates
           $[x_0,x_3]$ and $[x_1,x_5]$.
\end{itemize}

It is very interesting following the D-branes between these phases. In
phase 1 everything is smooth. This K3 surface contains a chain of 3
rational curves. The central curve is a conic and the two outside are
lines. Moving to phase 2 we contract the two lines to orbifold
singularities and moving to phase 3 we contract the conic to an
exoflop phase all as expected from section \ref{s:K3eg}. In phase 3 we
have one matrix factorization living at the North pole as shown in the
figure.

An interesting thing happens when we move from phase 1 to phase 3. We
bring the two lines together and they become a (singular)
conic. Accordingly, when we move to phase 5 these conics must induce
another exoflop. This pushes out the already exoflopped $\P^1$
producing a chain of two external $\P^1$'s.

On the other side of the pentagon in figure \ref{fig:face} we can contract
the conic in phase 2 to produce an exoflop, phase 4. This pushes 3
matrix factorizations onto the North pole. In this sense, phase 4
represents the noncommutative resolution of the $A_3$ singularity. If
we represent an $A_3$ singularity as $z_0^4-z_1z_2$ then the
noncommutative resolution using the methods in \cite{Aspinwall:2010mw}
is given by the quiver
\begin{equation}
\xymatrix@C=30mm@R=20mm{
  R \ar@<1mm>[r]^-{\left(\begin{smallmatrix}0\\1\end{smallmatrix}\right)}
  \ar@<1mm>[d]^-{\left(\begin{smallmatrix}1\\0\end{smallmatrix}\right)}
    &{\coker\begin{pmatrix}z_1&z_0\\z_0^3&z_2\end{pmatrix}}
  \ar@<1mm>[l]^-{\left(\begin{smallmatrix}z_2&-z_0\end{smallmatrix}\right)}
  \ar@<1mm>[d]^-{\left(\begin{smallmatrix}z_0&0\\0&1\end{smallmatrix}\right)}
\\
 {\coker\begin{pmatrix}z_1&z_0^3\\z_0&z_2\end{pmatrix}}
  \ar@<1mm>[u]^-{\left(\begin{smallmatrix}-z_0&z_1\end{smallmatrix}\right)}
  \ar@<1mm>[r]^-{\left(\begin{smallmatrix}1&0\\0&z_0\end{smallmatrix}\right)}
&{\coker\begin{pmatrix}z_1&z_0^2\\z_0^2&z_2\end{pmatrix}}
  \ar@<1mm>[u]^-{\left(\begin{smallmatrix}1&0\\0&z_0\end{smallmatrix}\right)}
  \ar@<1mm>[l]^-{\left(\begin{smallmatrix}z_0&0\\0&1\end{smallmatrix}\right)}
}
\end{equation}
The 3 matrix factorizations at the North pole in phase 4 correspond to
the 3 matrix factorizations in this quiver.

Passing from phase 4 to phase 5 is a new transition that we haven't
explored. The 3 matrix factorizations on the North pole in phase 4 are
``blown-up'' to produce the extra $\P^1$ in the double exoflop.

Let us note that the exoflop behaviour here occurs because the $A_3$
singularity manifests itself as a hypersurface singularity of the form
$z_0^4-z_1z_2$. If the singularity had appeared instead because of an orbifold
singularity in the ambient toric variety then we would have expected
orbifold phases rather than exoflop .

Generalizing this example we would expect an $A_{2k-1}$ singularity in
a K3 surface that presented itself as a hypersurface singularity 
$z_0^{2k}-z_1z_2$ to have a phase where a chain of $k$ $\P^1$'s
protrudes out from the singularity.


\section{Discussion}  \label{sec:conc}

A typical hypersurface (or complete intersection) \CY\ in a toric
variety has a huge number of phases and many of these phases involve
exoflops. This is true even for K3 surfaces as we have explored in
this paper. The exoflop seems to give a useful presentation of the
derived category (i.e., D-brane category) by manifestly ``splitting
off'' some objects by sending them to the North pole of an exoflopped
$\P^1$. While a semiorthogonal decomposition for a compact \CY\ is
impossible because of Serre duality, exoflops perhaps provide a useful
alternative in some sense.

We might regard something akin to phase 4 in figure \ref{fig:face}
where the matrix factorizations associated to the noncommutative
resolution all live at the North pole of a single $\P^1$ to be the
most useful picture, or it may be that the exoflop chain in phase 5
proves the best ``decomposition'' of the D-brane category. In
particular, it would be interesting, therefore, to have a general
understanding of the transitions of the type between phases 4 and 5.

The structure of exoflops is much richer for \CY\ threefolds than for
the case of dimension two we considered in this paper. More
importantly, exoflops are intimately connected to extremal transitions in
this case. The transitions that connect the web of \CY\ hypersurfaces
in toric varieties as studied in \cite{Avram:1997rs} all go via
exoflop limits. A flavour of this is seen in the K3 surface as
such transitions connect algebraic families of of K3's of different
generic Picard number. The example in section \ref{ss:conic} has
generic Picard number 2 but in the exoflop phase we have a quartic K3
surface. The analogue of an extremal transition would be to ignore the
exoflopped $\P^1$ and deform the complex structure of the remaining
part to a generic quartic with Picard number one.

This explicit exoflop picture of an extremal transition must
surely shed some light on the way the derived category changes as one
passes between topologically distinct \CY\ threefolds.

Another kind of phase is intimately connected to the exoflop
phase. This is the ``bad hybrid'' model of
\cite{me:hybridm,Bertolini:2013xga}. We can get a bad hybrid model
from figure \ref{fig:exoflop} by collapsing $X^\sharp$ to a point
using a \CY\ to Landau--Ginzburg transition. All that remains is a
$\P^1$ but it has two special points. At the North pole there is
$\O_C(-1)$ and at the South pole there are many matrix factorizations
that account for many the D-branes in $X^\sharp$. As in the exoflop
case, the grading prohibits the existence of any D-brane localized at
a point on the $\P^1$ except at the poles.

In a typical example with many phases, most phases would appear to
involve exoflops or bad hybrids. Given the equivalence of the D-brane
category between all the phases, it seems certain that these phases
are worthy of more attention.

\section*{Acknowledgments}

I thank N.~Addington and R.~Plesser for many useful and important
discussions. This work was partially supported by NSF grant
DMS--1207708.  Any opinions, findings, and conclusions or
recommendations expressed in this material are those of the author
and do not necessarily reflect the views of the National Science
Foundation.


\begin{thebibliography}{10}

\bibitem{W:phase}
E.~Witten,
\newblock {\em Phases of $N=2$ Theories in Two Dimensions},
\newblock Nucl. Phys. {\bf B403} (1993) 159--222, hep-th/9301042.

\bibitem{me:hybridm}
P.~S. Aspinwall and M.~R. Plesser,
\newblock {\em {Decompactifications and Massless D-Branes in Hybrid Models}},
\newblock JHEP {\bf 1007} (2010) 078, arXiv:0909.0252.

\bibitem{Bertolini:2013xga}
M.~Bertolini, I.~V. Melnikov, and M.~R. Plesser,
\newblock {\em {Hybrid Conformal Field Theories}},
\newblock JHEP {\bf 1405} (2014) 043, arXiv:1307.7063.

\bibitem{GMV:Ht}
B.~R. Greene, D.~R. Morrison, and C.~Vafa,
\newblock {\em A Geometric Realization of Confinement},
\newblock Nucl. Phys. {\bf B481} (1996) 513--538, hep-th/9608039.

\bibitem{AdAs:masscat}
N.~Addington and P.~S. Aspinwall,
\newblock {\em {Categories of Massless D-Branes and del Pezzo Surfaces}},
\newblock JHEP {\bf 1307} (2013) 176, arXiv:1305.5767.

\bibitem{DM:qiv}
M.~R. Douglas and G.~Moore,
\newblock {\em D-branes, Quivers, and ALE Instantons},
\newblock hep-th/9603167.

\bibitem{GKZ:book}
I.~M. Gelfand, M.~M. Kapranov, and A.~V. Zelevinski,
\newblock {\em Discriminants, Resultants and Multidimensional Determinants},
\newblock Birkh{\"a}user, 1994.

\bibitem{AGM:sd}
P.~S. Aspinwall, B.~R. Greene, and D.~R. Morrison,
\newblock {\em Measuring Small Distances in $N=2$ Sigma Models},
\newblock Nucl. Phys. {\bf B420} (1994) 184--242, hep-th/9311042.

\bibitem{me:enhg}
P.~S. Aspinwall,
\newblock {\em Enhanced Gauge Symmetries and K3 Surfaces},
\newblock Phys. Lett. {\bf B357} (1995) 329--334, hep-th/9507012.

\bibitem{CDFKM:I}
P.~Candelas et~al.,
\newblock {\em Mirror Symmetry for Two Parameter Models --- I},
\newblock Nucl. Phys. {\bf B416} (1994) 481--562, hep-th/9308083.

\bibitem{me:TASI-D}
P.~S. Aspinwall,
\newblock {\em D-Branes on {C}alabi--{Y}au Manifolds},
\newblock in J.~M. Maldacena, editor, ``Progress in String Theory. TASI 2003
  Lecture Notes'', pages 1--152, World Scientific, 2005,
\newblock hep-th/0403166.

\bibitem{MR2258045}
R.~Rouquier,
\newblock {\em Categorification of {${\frak{sl}}_2$} and braid groups},
\newblock in ``Trends in representation theory of algebras and related
  topics'', Contemp. Math.~{\bf 406}, pages 137--167, Amer. Math. Soc.,
  Providence, RI, 2006.

\bibitem{Anno:sph}
R.~Anno,
\newblock {\em Spherical Functors},
\newblock arXiv:0711.4409.

\bibitem{nick}
N.~Addington,
\newblock {\em New derived symmetries of some hyperk{\"a}hler varieties},
\newblock arXiv:1112.0487.

\bibitem{Horj:EZ}
R.~P. Horja,
\newblock {\em Derived Category Automorphisms from Mirror Symmetry},
\newblock math.AG/\-0103231.

\bibitem{ST:braid}
P.~Seidel and R.~P. Thomas,
\newblock {\em Braid Groups Actions on Derived Categories of Coherent Sheaves},
\newblock Duke Math. J. {\bf 108} (2001) 37--108, math.AG/0001043.

\bibitem{AD:Dstab}
P.~S. Aspinwall and M.~R. Douglas,
\newblock {\em D-Brane Stability and Monodromy},
\newblock JHEP {\bf 05} (2002) 031, hep-th/0110071.

\bibitem{Eis:mf}
D.~Eisenbud,
\newblock {\em Homological Algebra on a Complete Intersection, with an
  Application to Group Representations},
\newblock Trans. Amer. Math. Soc. {\bf 260} (1980) 35--64.

\bibitem{Orlov:mfc}
D.~Orlov,
\newblock {\em Derived Categories of Coherent Sheaves and Triangulated
  Categories of Singularities},
\newblock in ``Algebra, Arithmetic, and Geometry: in Honor of {Y}u. {I}.
  {M}anin. {V}ol. {II}'', Progr. Math.~{\bf 270}, pages 503--531, Birkh\"auser
  Boston Inc., Boston, MA, 2009,
\newblock math.AG/0503632.

\bibitem{HHP:linphase}
M.~Herbst, K.~Hori, and D.~Page,
\newblock {\em Phases Of $N=2$ Theories In $1+1$ Dimensions With Boundary},
\newblock arXiv:0803.2045.

\bibitem{me:csalg}
P.~S. Aspinwall,
\newblock {\em Topological D-Branes and Commutative Algebra},
\newblock hep-th/0703279,
\newblock submitted to Communications in Number Theory and Physics.

\bibitem{MR2795327}
E.~Segal,
\newblock {\em Equivalence Between {GIT} Quotients of {L}andau-{G}inzburg
  {B}-models},
\newblock Comm. Math. Phys. {\bf 304} (2011) 411--432.

\bibitem{AG:McExt}
L.~L. Avramov and D.~R. Grayson,
\newblock {\em Resulutions and Cohomology over Complete Intersections},
\newblock in D.~Eisenbud et~al., editors, ``Computations in Algebraic Geometry
  with Macaulay 2'', Algorithms and Computations in Mathematics~{\bf 8}, pages
  131--178, Springer-Verlag, 2001.

\bibitem{shipping}
D.~Halpern-Leistner and I.~Shipman,
\newblock {\em Autoequivalences of derived categories via geometric invariant
  theory},
\newblock arXiv:1303.5531.

\bibitem{MP:inst}
D.~R. Morrison and M.~R. Plesser,
\newblock {\em Summing the Instantons: Quantum Cohomology and Mirror Symmetry
  in Toric Varieties},
\newblock Nucl. Phys. {\bf B440} (1995) 279--354, hep-th/9412236.

\bibitem{bergh04:nc}
M.~Van~den Bergh,
\newblock {\em Non-Commutative Crepant Resolutions},
\newblock in ``The Legacy of Niels Henrik Abel: The Abel Bicentennial, Oslo
  2002'', pages 749--770, Springer, 2004,
\newblock arXiv:math/0211064.

\bibitem{Aspinwall:2010mw}
P.~S. Aspinwall and D.~R. Morrison,
\newblock {\em {Quivers from Matrix Factorizations}},
\newblock Commun. Math. Phys. {\bf 313} (2012) 607--633, arXiv:1005.1042.

\bibitem{Avram:1997rs}
A.~Avram, M.~Kreuzer, M.~Mandelberg, and H.~Skarke,
\newblock {\em {The Web of Calabi-Yau Hypersurfaces in Toric Varieties}},
\newblock Nucl.Phys. {\bf B505} (1997) 625--640, arXiv:hep-th/9703003.

\end{thebibliography}

\end{document}